\documentclass[12pt,
showpacs,showkeys,aps]{revtex4} 
\usepackage{amsmath}
\usepackage{amsfonts}
\usepackage{amssymb}
\usepackage[dvips]{graphicx}
\usepackage{epsfig}
\begin{document}

\title{Dynamical Casimir effect with Robin boundary conditions \\
in a three dimensional open cavity}

\author{C. Farina} \email{farina@if.ufrj.br}
\affiliation{Instituto de F\'{\i}sica, Universidade Federal do Rio
de Janeiro, Caixa Postal 68.528, 21941-972 Rio de Janeiro, RJ,
Brazil.}
\author{F. Pascoal}
\email{fabiopr@df.ufscar.br} \affiliation{Universidade Federal do
Rio de Janeiro, Campus de Macaé\\ Macaé, Rio de Janeiro, State
ZIP/Zone, Brasil.}
\author{D. Azevedo}\email{dazevedo@if.ufrj.br}
\affiliation{Instituto de F\'{\i}sica, Universidade Federal do Rio
de Janeiro, Caixa Postal 68.528, 21941-972 Rio de Janeiro, RJ,
Brazil.}

\begin{abstract}

We consider a massless scalar field in 1+1 dimensions
inside a cavity composed by a fixed plate, which imposes on
 the field a Robin BC, and an oscillating one, which
imposes on the field a Dirichlet BC. Assuming that
the  plate moves for a finite time interval, and
considering  parametric resonance, we compute the
total number of created particles inside the cavity. We
generalize our results to the case of two parallel plates in 3+1
dimensions.

\end{abstract}

\maketitle


\section{Introduction}\label{Introduction}

\indent

The dynamical Casimir effect (DCE) consists of two
related phenomena: real particle creation due to moving
boundaries and radiation reaction forces on moving boundaries.
 This effect already manifests itself for a unique moving plate and, for a non-relativistic motion,
the frequencies of the created particles (photons in the case of the
quantized electromagnetic field) are smaller or equal than the
mechanical frequency of the moving plate. Since Moore's pioneering
paper\cite{Moore70}, the DCE has been studied in many different
situations by many authors (for a review on this subject see \cite{DodonovReview} and
the special issue \cite{SpecialIssue}). Particularly, many distinct
boundary conditions (BC) have been considered, from the idealized
Dirichlet and Neumann ones to more realistic ones. However, the so
called Robin boundary conditions (RBC), which interpolate
continuously Dirichlet and Neumann ones, have rarely been used
explicitly in the context of the DCE (though they have been
considered by many authors in the context of the static Casimir
effect, see for instance \cite{Romeo-Saharian-2002}). As far as we know,
RBC appeared in the DCE only for the situation of
one  moving plate in 1+1 dimensions \cite{Bruno-1,Bruno-2}. Our
purpose here is to consider RBC in  one-dimensional cavities with one oscillating wall and
in three-dimensional (open) cavities formed by two parallel
plates with one of them oscillating in time.
For a scalar field $\phi$ in 3+1 dimensions, RBC are defined by
$
\phi\vert_{bound.}= \beta \frac{\partial \phi}{\partial n}  \vert_{bound.}
$
where  $\beta$ is a constant parameter with dimension of length. They interpolate continuously Dirichlet ({$\beta \to 0$}) and
Neumann ({$\beta \to \infty$}) BC. They appear in different areas of physics:
 from Mechanics, Electromagnetism and Quantum Mechanics to Quantum Field Theory, among others.
These BC were used as a phenomenological model for penetrable surfaces \cite{Mostepanenko-Trunov-1985}.
In fact, for  {$\omega\ll\omega_P$}, parameter $\beta$ plays the role of the plasma wavelength.
In Classical Mechanics, RBC may appear in a vibrating string coupled to a harmonic oscillator at one of its edges
   \cite{Chen-Zhou-1992,Bruno-1}.
In the context of the static Casimir effect, RBC lead to  eigenfrequencies for the cavity modes that are roots of a transcendental equation.
 %
%

In the context of the Dynamical Casimir effect,  Mintz {\it et al} \cite{Bruno-1} considered
 a massless scalar field $\phi$ in $1+1$ dimensions under the influence of one moving boundary in a
  prescribed and  non-relativistic motion with small amplitudes, namely,
$
|\delta \dot{q}(t)|<<c$  and $|\delta q(t)|<<c/\omega_0,
$
where $\delta q(t)$ is the position of the moving boundary at instant $t$ and $\omega_0$ is the
dominant mechanical frequency.
%
Using the perturbative approach of Ford and Vilenkin \cite{Ford-Vilenkin-1982}, 
 the solution of the wave equation, $\partial^2\phi (t,x)=0$, submitted to a RBC, leads
to a susceptibility with both real and imaginary parts, so that,
$
\delta{\mathcal F}(\omega) = {\chi(\omega)} \delta Q(\omega)
$,
with
$
\chi(\omega) = {\mathcal R}e\chi(\omega) + i{\mathcal I}m\chi(\omega).
$
Recall that, for the same situation, the use of a Dirichlet (or Neumann) BC would lead
to a purely imaginary susceptibility $\chi\!_{{\,}_D}(\omega)=\omega^3/6\pi$,  (${c=\hbar=1}$).
For a typical oscillatory motion, given by
$
 \delta q(t) = \delta q_0 \, e^{-\vert t\vert/T}\!\!\cos(\omega_0 t),
 $
 with
 \linebreak
 $\omega_0T\gg 1,
$
 Mintz {\it et al} \cite{Bruno-1} showed that the dissipative force on the moving boundary can
be enormously suppressed for  $\beta \omega_0 \approx 2$. In a subsequent paper \cite{Bruno-2}, these authors analyzed
 the particle creation phenomenon for the same situation and found that, for the above mentioned relation between $\beta$ and $\omega_0$,
 there is also an enormous suppression of particle creation.
 %

\section{One-dimensional cavities with Robin BC}

\indent

For simplicity, we consider a one-dimensional cavity composed by a fixed plate at $x=0$, which imposes
on the massless scalar field a RBC, and a moving plate whose position at instant $t$ is given by $q(t)$, which imposes on
the field a Dirichlet BC. Then, we must
solve the wave equation with $\hat\phi$ submitted to
%
%
$
\left( \hat\phi(x,t) -\gamma \frac{\partial}{\partial x}\hat\phi(x,t)\right)
\big|_{x=0} \!=\! 0\;
  \mbox{and}\;
   \hat\phi(x,t) \big|_{x=q(t)} \!= \!0.
   $
   %
 %
The scalar field satisfies the usual field commutation relations, namely,
$
\left[ \hat{\phi}(x;t),\hat{\pi}(x^{\prime };t)\right] = i\delta
(x^{\prime }-x),
$
and
$
\left[ \hat{\phi}(x;t),\hat{\phi}(x^{\prime };t)\right] =
\left[ \hat{\pi}(x;t),\hat{\pi}(x^{\prime };t)\right] =0,
$
Our anstaz for the field operators
$\hat\phi$ and $\hat\pi = \dot{\hat\phi}$ in terms of the instantaneous basis introduced by Law\cite{Law-1994}, in 1994,  is
       %
          %
          \begin{eqnarray}
            \hat{\phi}(x;t) &=& \sum_{n}\frac{1}{\sqrt{2k_{n}(t)}}\;u_{n}(x;t)
            \Bigl\{\hat{\textsl{a}}_{n}(t)+\hat{\textsl{a}}_{n}^{\dag }(t)\Bigr\}  \label{Campo-phi}
            \\
            \hat{\pi}(x;t) &=&-i\sum_{n}\sqrt{
            \frac{k_{n}(t)}{2}}\;u_{n}(x;t)
            \Bigl\{\hat{\textsl{a}}_{n}(t)-\hat{\textsl{a}}_{n}^{\dag }(t)\Bigr\} ,
            \label{Campo-pi}
          \end{eqnarray}
%
where the modes $\{ u_n(x,t)\}$ of the instantaneous basis must satisfy Helmholtz equation,
$
\left( \frac{\partial^{2}}{\partial x^{2}}+k_{n}^{2}(t)\right) u_{n}(x;t)=0,
$
the boundary conditions
$
\left(u_n(x,t) - \gamma\frac{\partial \ }{\partial x}
 u_n(x,t)\right)\vert_{x=0} = 0
 $
 and
$
  u_n(x,t)\vert_{x=q(t)} = 0\, ,
$
and the orthonormality condition $\int_{0}^{q(t)}\!\!\mbox{d}x u_{n}(x;t) u_{m}(x;t) \!= \!\delta _{nm}$.
With these properties, it follows that
$
\left[ \textsl{a}_{n}(t),\textsl{a}_{m}^{\dag }(t)\right] = \delta _{nm}
$
and
$
\left[ \textsl{a}_{n}^{\dag }(t),\textsl{a}_{m}^{\dag }(t)\right]=
\left[ \textsl{a}_{n}(t),\textsl{a}_{m}(t)\right] =0.
$
The instantaneous basis can be explicitly obtained, with modes {$u_n(x,t)$} given by
\begin{equation}\label{ModoBaseInstantanea}
u_{n}(x;t)=\frac{A_{n}(t)}{\sqrt{2 q(t)}}\sin\Bigl[ k_n(t) (x-q(t)) \Bigr],
\end{equation}
where
$
%
A_{n}(t)=2\left[ 1+\frac{\gamma /q(t)}{1+\gamma ^{2}
 k_{n}^{2}(t)}\right] ^{-1/2}
 $
%
and $\{k_n(t)\}$ are the roots of the following transcendental equation
%
$
\sin[q(t) k_{n}(t)] + \gamma k_{n}(t)
 \cos \bigl[q(t) k_{n}(t)\bigr] = 0 .
 $

\begin{figure}[!h]
  \begin{center}
    \includegraphics[scale=0.75]{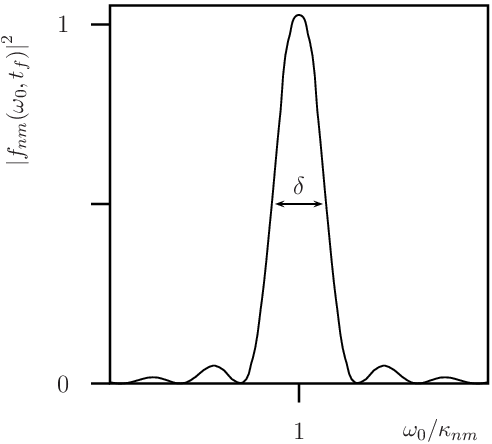}
    \caption{\footnotesize  {$\vert f_{nm}(\omega_0;t_f)\vert^2$} as a function of {$\omega_0/\kappa_{nm}$}  for
    {$\omega_0 t_f\gg1$}. }
    %
    \label{Function-fmn}
  \end{center}
\end{figure}

 Time evolution equations for  ${\hat a}_n(t)$ and ${\hat a}^\dagger_n(t)$ can be found,
        \begin{equation}\label{EqDif-an-Xi-Lambda}
          \dot{\hat{\textsl{a}}}_{n}(t) = -ik_{n}(t)\hat{\textsl{a}}_{n}(t) +
          \sum_j \Xi_{jn}(t)  \hat{\textsl{a}}_{j}(t) +
          \sum_j \Lambda_{jn}(t) \hat{\textsl{a}}_{j}^{\dag }(t) \, ,
        \end{equation}
  where
\begin{eqnarray}\label{Xi}
\Xi _{mn}(t) &:=&-\frac{1}{2}G_{mn}(t)\left( \sqrt{\frac{k_{n}(t)}{k_{m}(t)}}+%
\sqrt{\frac{k_{m}(t)}{k_{n}(t)}}\right) ;  \\
\Lambda _{mn}(t) &:=&\frac{\dot{k}_{n}(t)}{2k_{n}(t)}\delta _{mn}-\frac{1}{2}%
  G_{mn}(t)\left( \sqrt{\frac{k_{n}(t)}{k_{m}(t)}}-\sqrt{\frac{k_{m}(t)}{%
  k_{n}(t)}}\right)  \label{Lambda}\, ,
  \end{eqnarray}
with
%
 $
  G_{nm}(t):=\int_{0}^{q(t)}\mbox{d}x\ \dot{u}_{n}(x;t)u_{m}(x;t)
  $
  (an analogous equation holds for ${\hat a}^\dagger_n(t)$)).
%
 Relating $ \hat{\textsl{a}}_{n}$ and $\hat{\textsl{a}}_{n}^{\dag}$ for different times, we write
\begin{equation}\label{CoefBogoliubov}
\hat{\textsl{a}}_{n}(t)=\sum_{m}\alpha_{nm}(t)\hat{\textsl{a}}_{m}(t_{0}) +
\sum_{m}\beta_{nm}(t)\hat{\textsl{a}}_{m}^{\dag}(t_{0})\, ,
\end{equation}
where the Bogoliubov coefficients must satisfy $\alpha_{nm}(t_{0})=\delta_{nm}$ and $\beta_{nm}(t_{0})=0$.
 The time evolution of these coefficients can be established,
\begin{eqnarray}\label{AlphamnPonto}
\dot{\alpha}_{nm}(t) \!&=&\!-ik_{n}(t)\alpha _{nm}(t) +
\sum_{j}\Xi _{jn}(t)\alpha_{jm}(t)+\sum_{j}\Lambda _{jn}(t)\beta _{jm}^{\ast }(t); \\
\dot{\beta} _{nm}(t) \!&=&\!-ik_{n}(t)\beta _{nm}(t)+\sum_{j}\Xi _{jn}(t)\beta
_{jm}(t)+\sum_{j}\Lambda _{jn}(t)\alpha _{jm}^{\ast}(t).\label{BetamnPonto}
\end{eqnarray}
Previous equations may be simplified with the aid of definitions:
%
          %
          \begin{eqnarray}\label{AlphamnTilde}
            \alpha _{nm}(t) &=:&\mbox{e}^{-iK_{n}(t)}\tilde{\alpha}_{nm}(t)\, ;
            \;\;\;\;
            \beta _{nm}(t) =:\mbox{e}^{-iK_{n}(t)}\tilde{\beta}_{nm}(t)\, ;
            \;\;\;\;
             \cr
            K_{n}(t) &:=&   \int_{t_{0}}^{t}\mbox{d}t^{\prime}\ k_{n}(t^{\prime });\\
          %
            \Xi _{mn}(t) &=:& \tilde{\Xi}_{mn}(t)\mbox{e}^{i\left[
            K_{m}(t)-K_{n}(t)\right] }\, ;\;\;
            \Lambda_{mn}(t) =: \tilde{\Lambda}_{mn}(t)
            \mbox{e}^{-i\left[ K_{m}(t)+K_{n}(t)\right] }\;,
            \nonumber\label{LambdamnTilde}
          \end{eqnarray}
          %
    %
   Consequently, the time evolution for coefficients ${\tilde{\alpha}}_{nm}$ and ${\tilde{\beta}}_{nm}$ are
          %
          %
          {
     \begin{eqnarray}\label{AlphamnTildePonto}
           \dot{\tilde{\alpha}}_{nm}(t) &=&\sum_{j}\tilde{\Xi}_{jn}(t)\tilde{\alpha}
           _{jm}(t)+\sum_{j}\tilde{\Lambda}_{jn}(t)\tilde{\beta}_{jm}^{\ast }(t);
          \\
           \dot{\tilde{\beta}}_{nm}(t) &=&
          \sum_{j}\tilde{\Xi}_{jn}(t)\tilde{\beta}_{jm}(t) +
           \sum_{j}\tilde{\Lambda}_{jn}(t)
          \tilde{\alpha}_{jm}^{\ast }(t) \; .\label{BetamnTildePonto}
    \end{eqnarray}
    }
%
 Up to this point, our calculations are exact. However, from now on, we shall consider only
 oscillating motions with small amplitudes, so we write
$
q(t)=q_0[1 + {\epsilon} \xi(t)]$, with ${\epsilon\ll 1}
$
and $\xi(t)$ given, for a typical motion, by
          \begin{align*}
              \xi(t) =
              \begin{cases}
                    \sin(\omega_0 t) & 0<t<t_f \\
                    0 & t\le 0 \text{ or } t\ge t_f \;.
              \end{cases}
          \end{align*}
Expansions in powers of $\epsilon$ (recall that all quantities get an implicit
 \linebreak
 $\epsilon$-dependence through $q(t)$) lead to
 %
          \begin{eqnarray}\label{AlphanmTildeFinal}
            \tilde{\alpha}_{nm}^{(\ell )}(t)\! &=&\!\sum_{\ell ^{\prime }=1}^{\ell
            }\sum_{j}\int_{t_{0}}^{t}\mbox{d}\tau \left[ \tilde{\Xi}_{jn}^{(\ell
            ^{\prime })}(\tau )\tilde{\alpha}_{jm}^{(\ell -\ell ^{\prime })}(\tau
            )+\tilde{\Lambda}_{jn}^{(\ell ^{\prime })}(\tau )\tilde{\beta}
            _{jm}^{\ast (\ell -\ell ^{\prime }) }(\tau )\right] ; \\
             \tilde{ \beta}_{nm}^{(\ell )}(t)
             \!&=&\!
             \sum_{\ell ^{\prime }=1}^{\ell
            }\sum_{j}\int_{t_{0}}^{t}\mbox{d}\tau \left[ \tilde{\Xi}_{jn}^{(\ell
            ^{\prime })}(\tau )\tilde{\beta}_{jm}^{(\ell -\ell ^{\prime })}(\tau )+
            \tilde{\Lambda}_{jn}^{(\ell ^{\prime })}(\tau )\tilde{\alpha}
            _{jm}^{\ast (\ell -\ell ^{\prime }) }(\tau )\right]\, .
            \label{BetanmTildeFinal}
          \end{eqnarray}
  where the superscritps mean the order of the derivative respect to $\epsilon$ of the quantity in question and
       conditons $\tilde{\alpha} _{nm}^{(0)}(t)=\delta _{nm}$ and $\tilde{\beta}_{nm}^{(0)}(t)=0$ are satisfied.
The number of particles created inside the cavity, with energy $\omega_n=k_n$, after the motion is finished is given by
%
          \begin{equation}\label{eq:3.24}
            \mathcal{N}_{n}(t_f) = \left\langle 0\right\vert \hat{\textsl{a}}_{n}^{\dag }
            \left( t_{f}\right)\hat{\textsl{a}}_{n}
            \left( t_{f}\right) \left\vert 0\right\rangle =\sum_{j}\left\vert
            \sum_{\ell} \epsilon^{\ell} {\beta}^{(\ell)}
            _{nj}(t_{f})\right\vert ^{2}.
          \end{equation}
          %
\noindent
The first correction to $\mathcal{N}_{n}(t_f)$ occurs at order $\epsilon^2$,
  %
  \begin{equation}
  \mathcal{N}_{n}(t_f) = \epsilon ^{2}\sum_{m}\left\vert
  \beta_{nm}^{(1)}(t_{f})\right\vert ^{2} =
  \epsilon ^{2}\sum_{m}\left\vert
  \tilde\beta_{nm}^{(1)}(t_{f})\right\vert ^{2}\, ,
  \end{equation}
  %
    \noindent
  For the motion in consideration, we have
%
\begin{equation}\label{Ntf1+1}
\mathcal{N}_{n}(t_f) = \sum_{m}\left\vert \mathcal{C}_{nm}(\gamma)f_{nm}(\omega_0
,t_{f})\right\vert ^{2}\left( \epsilon \omega_0 t_{f}\right) ^{2}.
\end{equation}
%
\noindent
where
%
          \begin{eqnarray}\label{Cnm(gamma)}
          %
              f_{nm}(\omega_0;t) &{:=}& \frac{e^{i(\omega_0 + \kappa_{nm})t} \; -\; 1}
              {(\omega_0 + \kappa_{nm})t\;\;}\; -\;
              \frac{e^{-i(\omega_0 - \kappa_{nm})t} \; -\; 1}
              {(\omega_0 - \kappa_{nm})t\;\;}\cr
                 C_{nm}(\gamma) &=& \frac{1}{8} A_n(0)A_m(0)
            \frac{\sqrt{k_n(0)k_m(0)}}{k_n(0) + k_m(0)}\; ,\cr
               \kappa_{nm} &=& k_n(0) + k_m(0)\, .\nonumber
          \end{eqnarray}
At this order, the total number of particles created inside the cavity is given by
$
\mathcal{N} = \sum\!\!\!\!\!\!\!\!\!_{{\,}\atop{{\,}_{n,m}}}\left\vert \mathcal{C}_{nm}(\gamma)f_{nm}(\omega_0
,t_{f})\right\vert ^{2}\left( \epsilon \omega_0 t_{f}\right)^{2}
$
while the total energy of the created particles is given by
$
\mathcal{E} = \sum\!\!\!\!\!\!\!\!\!_{{\,}\atop{{\,}_{n,m}}}k_{n}\left\vert \mathcal{C}_{nm}(\gamma)
f_{nm}(\omega_0 ,t_{f})\right\vert ^{2}\left( \epsilon \omega_0 t_{f}\right)^{2}.
$
The behavior of $\vert f_{nm}(\omega_0;t_f)\vert^2$ is shown in Figure \ref{Function-fmn}.
For $\omega_0 t_f\gg1$, it has a peak around  $\omega_0=\kappa_{mn}$ whose width $\delta$ is proportional to
$1/(\kappa_{nm}t_f)$ (a simple estimative gives $\delta \approx 5.6/(\kappa_{nm}t_f)$). Hence,
   in a first approximation, $\vert f_{nm}(\omega_0;t_f)\vert^2$ behaves like a delta function,
showing that whenever the oscillation frequency $\omega_0$ equals
the sum of two energy levels of the corresponding static cavity we
have the best conditions for particle creation.
 \vskip 0.3 cm

%

\begin{figure}[!hbt]
\begin{center}
\includegraphics[width=0.52\textwidth]{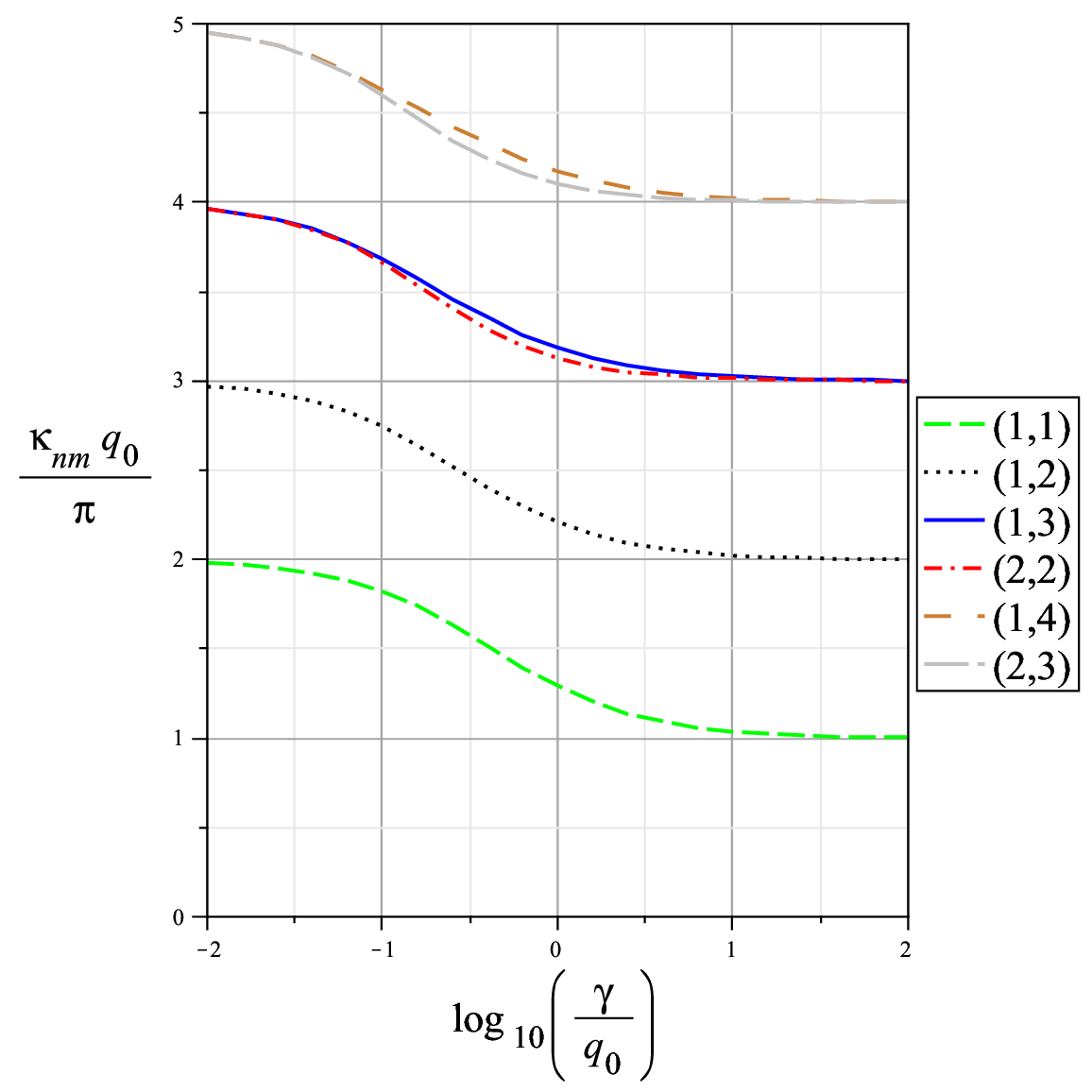}
 \caption{\footnotesize Resonances (in units of $\pi/q_0$) as functions of $\log_{10}(\gamma/q_0)$.}
 \label{KappaVersusLogGamma}
\end{center}
\end{figure}

 The fact that $\kappa_{mn}$ is given by
a sum of 2 terms means that particles are created in pairs.
The set of values of $\kappa_{mn}$ are called the resonances of the problem. Note that, for each
value of the Robin parameter, $\gamma$, we have a different set of resonances. Figure \ref{KappaVersusLogGamma} shows
how the resonances vary with  $\gamma$. Since $\gamma$ varies from $0$ (Dirichlet BC) to $\infty$ (Neumann BC),
it is convenient to make the plot against $log_{10}(\gamma/q_0)$, instead of $\gamma$.
%

%

For a given value of $\gamma$, the  resonances are obtained  by tracing a vertical line
and looking at the intersections in Figure \ref{KappaVersusLogGamma}. The values obtained this way for $\log_{10}(\gamma/q_0) = -2$ (extreme left on the graph) are, approximately, the resonances for Dirichlet-Dirichlet BC since,  for this case, $\gamma\ll q_0$. By the same token, the values obtained this way for $\log_{10}(\gamma/q_0) = 2$ (extreme right on the graph) are, approximately,
the resonances for Neumann-Dirichlet BC since, for this case, $\gamma\gg q_0$.
 Adjacent resonances are equally spaced only for D-D and N-D cases. For these cases we have degeneracies, which are broken
 in the Robin-Dirichlet case. For instance, for this last case, $\kappa_{13}= k_1 + k_3 \ne k_2 + k_2 = \kappa_{22}$, as can be
  seen in Figure \ref{KappaVersusLogGamma} near $\log_{10}(\gamma/q_0) = 0$. Note, also, the monotonic
 behavior of the curves with $\gamma/q_0$.

\begin{figure}[!hbt]
\begin{center}
\includegraphics[width=0.49\textwidth]{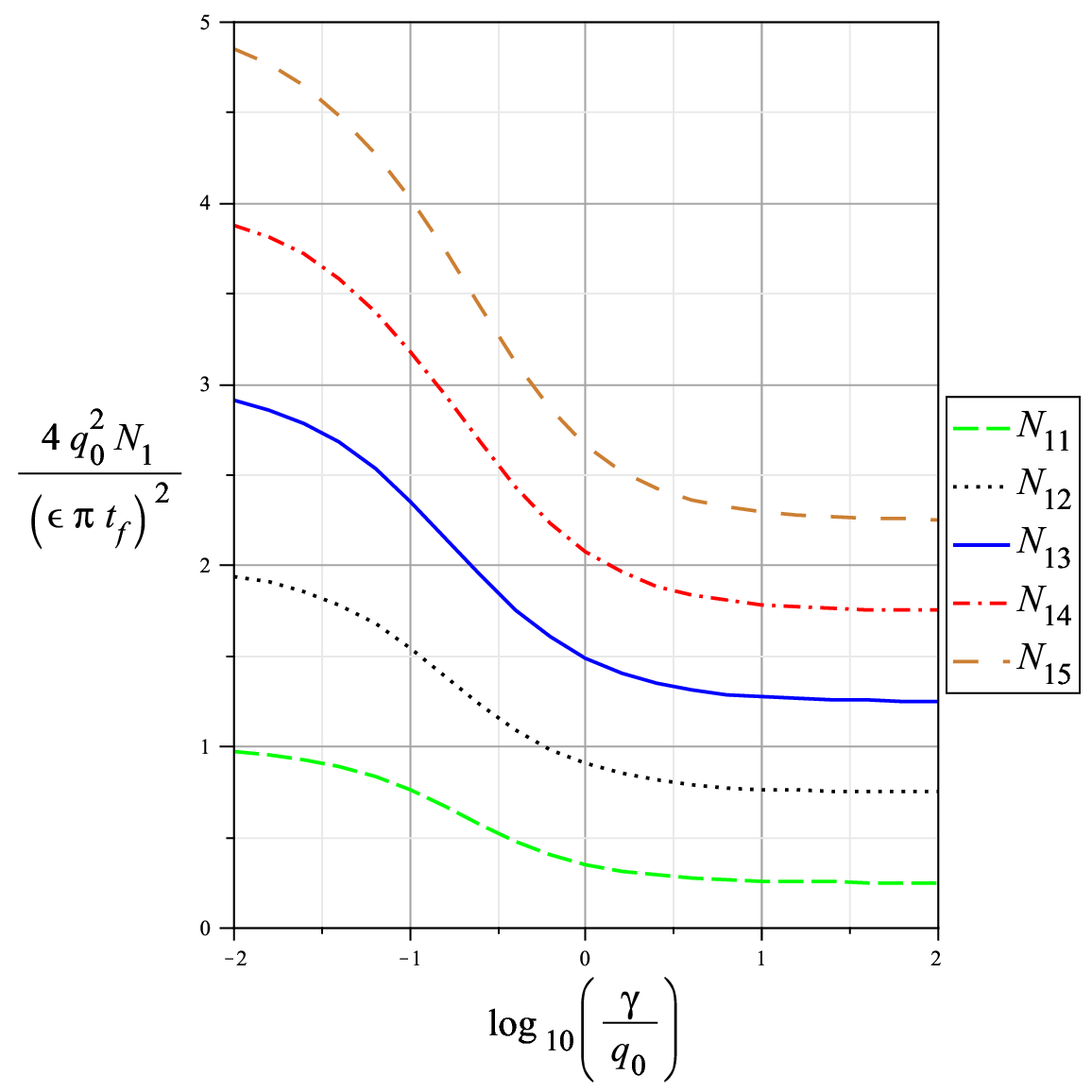}
\caption{\footnotesize Number of created particles with energy $k_1(0)$ (in units of
 $\left(\frac{\epsilon\pi t_f}{2q_0}\right)^2$ for the first five
resonant values of $\omega_0$p as a function of $\log_{10}(\gamma/q_0)$.}
\label{N1VersusLogGamma}
\end{center}
\end{figure}

Figure \ref{N1VersusLogGamma} shows the number of created particles with energy $k_1$  for different resonant values of the
 mechanical frequency as a function of $\log_{10}(\gamma/q_0) $ (since particles are created in pairs, 
 there are many ways of creating particles with energy $k_1$, namely, $\omega_0=\kappa_{11}$, $\omega_0=\kappa_{12}$, etc.).
 For the resonance $\omega_0=\kappa_{1m}$, we have  $N_{1m} = \mathcal{C}^2_{1m}(\gamma )\left( \epsilon \kappa_{1m}
  t_{f}\right) ^{2}$.
%

 %

It is worth saying a few  words about how the  curves in Figure \ref{N1VersusLogGamma} are traced. For each value of $\gamma$, we compute numerically
the set of corresponding resonances. Then, we compute $N_{11}$,  $N_{12}$,  $N_{13}$, ..., for that value of $\gamma$. We, then,
take another value of $\gamma$ and compute the new values of the resonances. Taking  $\omega_0$ equal to the new values of resonances we
compute again $N_{11}$,  $N_{12}$,  $N_{13}$, and so on. Hence,  distinct points of a given curve, for instance  $N_{1}$,  are
computed with distinct values of $\omega_0$, but with $\omega_0$ always equal to the first
resonance ($\kappa_{11}$, which depends on $\gamma$). Note, also, the monotonic behavior of curves in Figure \ref{N1VersusLogGamma}.

Let us check some particular cases. For $\omega_0 = \frac{2\pi }{q_0}$ and  $\gamma =0$, which corresponds
to the D-D case with $\omega_0=2k_1$ (parametric resonance with the lowest level of the  static cavity), we have
$
\mathcal{N}_{1}\simeq \left( \frac{\epsilon \pi t_{f}}{ 2 q_0}\right) ^{2}
$
and
$
\mathcal{E}_1 \simeq \frac{\pi}{q_0}\left( \frac{\epsilon \pi t_{f}}{2 q_0} \right) ^{2}
$
in agreement with  Dodonov and Klimov \cite{Dodonov-Klimov-1996}.
For $\omega = \frac{\pi}{q_0}$ and $\gamma \rightarrow\infty$, which corresponds to the N-D case with
$\omega_0=2k_1$ (parametric resonance with the lowest level of the  static cavity, which is $\frac{1}{2}$ the value for
the D-D case), we have
$
\mathcal{N}_{1}\simeq \left( \frac{\epsilon \pi t_{f}}{ 4 q_0}\right) ^{2}
$
and
$
\mathcal{E}_1\simeq \frac{\pi}{2 q_0}\left( \frac{\epsilon \pi t_{f}}{4 q_0} \right) ^{2},
$
in agreement with  Alves {\it et al} \cite{AlvesEtAl-2006}.
%
%

%
\section{Parallel plates in 3+1 dimensions with Robin BC}

\begin{figure}[!hbt]
 \begin{center}
 \includegraphics[width=0.95\textwidth]{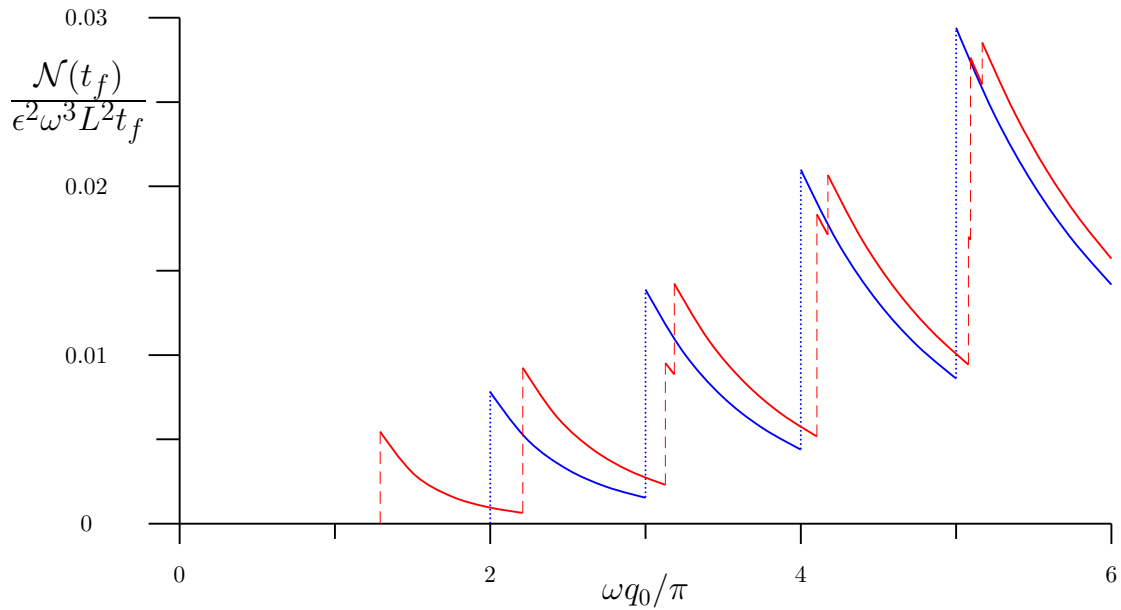}
 \caption{\footnotesize Total number of created particles for an open three-dimensional cavity formed by two parallel plates
 as a function of the frequency of the moving plate.}
 \label{ParticulasCriadasPlacas3D}
 \end{center}
 \end{figure}

Here, we shall generalize  some of the previous results to 3+1 dimensions. We, then, consider a fixed plate at
$z=0$, which imposes on a massless scalar field a RBC and a moving plate, parallel to the first one, which imposes on the field
a DBC. Let  $q(t)$ be the position of the moving plate at instant $t$. Operators
$\hat{\phi}\left( \mathbf{x};t\right)$ and $\hat{\pi}\left( \mathbf{x};t\right)$ are given, in terms of instantaneous basis, by
%
%
\begin{eqnarray}
\hat{\phi}\left( \mathbf{x};t\right) \!&=&\!\sum_{n=1}^{\infty }\int \frac{%
\mbox{d}^{2}\mathbf{k}_{\Vert }}{\sqrt{2\omega _{n}\left( k_{\Vert
},t\right) }}~u_{n}\left( x;t\right) \left[ \frac{\mbox{e}^{i\mathbf{k}%
_{\Vert }\cdot \mathbf{x}}}{2\pi }\hat{a}_{n}\left(
\mathbf{k}_{\Vert },t\right) +\mbox{h.c.}\right] , \\
\hat{\pi}\left( \mathbf{x};t\right)\!\! &=& \!\!-i\sum_{n=1}^{\infty }\!\!\int
\!\! \mbox{d}^{2}\mathbf{k}%
_{\Vert }~\!\!\sqrt{%
\frac{\omega _{n}\left( k_{\Vert },t\right) }{2}}~\!\!u_{n}\left( x;t\right) \!\!\left[
\frac{\mbox{e}^{i\mathbf{k}_{\Vert
}\cdot \mathbf{x}}}{2\pi }\hat{a}_{n}\left( \mathbf{k}_{\Vert },t\right) -%
\mbox{h.c.}\!\right]
\end{eqnarray}
where $\omega _{n}^{2}\left( k_{\Vert },t\right)\! = \!k_{\Vert }^{2}+k_{n}^{2}(t)$
 and
$
u_{n}\left( x,t\right)\! = \!\!\sqrt{\frac{2}{q(t)}}{A_{n}(t)}\sin \bigl[
{k_{n}(t)}\left( x - q(t)\right) \bigr] ,
$
with {$A_{n}(t)$} and {$k_{n}(t)$} defined as in the  1+1 case. We shall consider the same
motion as in the 1+1 case. The Bogoliubov coefficients are now defined by
%
\begin{equation}
\hat{a}_{n}\left( \mathbf{k}_{\Vert },t\right) =\sum_{m=0}^{\infty
}\left[ \alpha _{nm}\left( k_{\Vert },t\right) \hat{a}_{m}\left(
\mathbf{k}_{\Vert
},0\right) +\beta _{nm}\left( k_{\Vert },t\right) \hat{a}_{m}^{\dag }\left( -%
\mathbf{k}_{\Vert },0\right) \right] .  \label{CB}
\end{equation}
%
A perturbative solution, up to first order in $\epsilon $, leads to
%
\begin{equation}
\beta _{nm}\left( k_{\Vert },t\right) \mbox{e}^{i\omega _{n}\left(
k_{\Vert }\right) t} = -\epsilon\, C_{nm}\left( k_{\Vert
}\right) f_{nm}\left( k_{\Vert },t\right) ,
\end{equation}
where we defined
%
%
\begin{eqnarray}
C_{nm}\left(
k_{\Vert }\right)  &=&\frac{A_{n}(0)A_{m}(0)}{\sqrt{\omega
_{m}\left( k_{\Vert }\right) \omega _{n}\left( k_{\Vert }\right) }}\frac{%
k_{n}(0)k_{m}(0)}{\omega _{n}\left( k_{\Vert }\right) +\omega _{m}\left(
k_{\Vert }\right) },\\
f_{nm}\left( k_{\Vert },t\right)  &=& \int_{0}^{t}\mbox{d}t^{\prime }
\dot{\xi}(t^{\prime })\mbox{e}^{i\left[ \omega _{m}\left( k_{\Vert
}\right) +\omega _{n}\left( k_{\Vert }\right) \right] t^{\prime }}
\end{eqnarray}
with {$\omega _{n}^{2}\left( k_{\Vert }\right) =\omega _{n}^{2}\left(
k_{\Vert },0\right) $}. The number of created particles in a given mode with $k_z = k_n$ and with a parallel moment between
$\mathbf{k}_{\Vert }$ and $\mathbf{k}_{\Vert }+$d$^{2}\mathbf{k}_{\Vert }$ is
%
\begin{equation}
\mathcal{N}_{n}\left( \mathbf{k}_{\Vert },t_{f}\right) \mbox{d}^{2}\mathbf{k}
_{\Vert }
 =  \epsilon ^{2}\frac{L^{2}}{\left(
2\pi \right) ^{2}}\sum_{m=1}^{\infty }\left\vert C_{nm}\left( k_{\Vert
}\right) f_{nm}\left( k_{\Vert },t_{f}\right) \right\vert ^{2}%
\mbox{d}^{2}\mathbf{k}_{\Vert }.
\end{equation}
%
\noindent
The total number of created particles inside the cavity takes the form
%
\begin{equation}
\mathcal{N}\left( t_{f}\right)
 = \epsilon ^{2}\frac{L^{2}
}{2\pi }\sum_{n,m=1}^{\infty }\int_{0}^{\infty }\mbox{d}k_{\Vert
}~k_{\Vert }\left\vert C_{nm}\left( k_{\Vert }\right)
f_{nm}\left( k_{\Vert },t_{f}\right) \right\vert ^{2},
\end{equation}
%
\noindent
and the total energy is given by
%
\begin{equation}
\mathcal{E}\left( t_{f}\right)
=\epsilon ^{2}\frac{L^{2}}{2\pi
}\sum_{n,m=1}^{\infty }\int_{0}^{\infty }\mbox{d}k_{\Vert }~k_{\Vert
}\omega _{m}\left( k_{\Vert }\right) \left\vert C_{nm}
\left( k_{\Vert }\right) f_{nm}\left( k_{\Vert
},t_{f}\right) \right\vert ^{2}.
\end{equation}
%
%
For the harmonic motion considered before, with $\omega t_{f}\gg1$), we get
%
\begin{eqnarray}
\left\vert f_{nm}\left( k_{\Vert },t_{f}\right) \right\vert
^{2}
&=&
\frac{\pi \omega ^{2}t_{f}}{4k_{\Vert
}}\left( \frac{\omega _{n}\left( k_{\Vert }\right) \omega _{m}\left(
k_{\Vert }\right) }{\omega _{n}\left( k_{\Vert }\right) +\omega
_{m}\left( k_{\Vert }\right) }\right) \delta \left( k_{nm}(\omega
)-k_{\Vert }\right) ,
\end{eqnarray}
where
$
\sqrt{k_{nm}^{2}(\omega )+k_{n}^{2}}+\sqrt{k_{nm}^{2}(\omega )+k_{m}^{2}}%
-\omega =0.
$
Using last result for $\left\vert f_{nm}\left( k_{\Vert
},t_{f}\right)\right\vert ^{2}$, we obtain
%
\begin{equation}\label{Ntf3+1}
\mathcal{N}\left( t_{f}\right) = \epsilon ^{2}\frac{L^{2}t_{f}}{8\omega }
\sum_{n,m=1}^{\infty }\left( A_{n}A_{m}k_{n}k_{m}\right) ^{2}\Theta
\left( \omega-k_{n}-k_{m}\right)
\end{equation}
and $\mathcal{E}\left( t_{f}\right) = \mathcal{N}\left( t_{f}\right) \omega/2$.
Figure \ref{ParticulasCriadasPlacas3D} shows the behavior of the total number of created particles inside the plates in terms of
the frequency $\omega$ of the moving plate. We plot $\mathcal{N}\left( t_{f}\right)$ divided
by $\epsilon^2\omega^3 L^2 t_f$ in terms of $\omega a_0/\pi$.
%

%
%

Solid lines  connected by dotted lines correspond to the DD case, while solid lines
 connected by dashed lines, to a RD case. The discontinuities occur at the resonant values
($\omega = k_n + k_m$). The main difference between DD and RD cases consists in the fact that the resonances for the former are
equally spaced, while for the latter they are not equally spaced, as can be seen from Figure \ref{ParticulasCriadasPlacas3D}.
Note the presence of small solid lines for the RD case, a direct consequence of the degeneracy breaking that happens
when we use RBC, as discussed previously. It is worth noting the similarity of the graph for the D-D case with that for the electromagnetic field inside two parallel and perfectly conducting plates discussed by Mundarain and Maia Neto\cite{Mundarain-PAMN-1998}.

\pagebreak
\section{Final comments}

\indent


In this work we considered RBC in one-dimensional cavities and in a
three-dimensional  open cavity formed by two parallel plates.  Using
the instantaneous basis method \cite{Law-1994} we computed the
number of created particles when the  frequency of the oscillating
plate was at resonance. As we showed, for one-dimensional cavities,
there are more resonances for the RD case than for the DD or ND
cases, due to the degeneracy breaking discussed in the text. For the
same reason,  there are more discontinuities in Figure
\ref{ParticulasCriadasPlacas3D}
 when a RBC is involved than for the case where both plates impose
 a DBC on the field. An important difference between the 1+1 and 3+1 cases
 treated here is that in the former the total number of created particles,
 $\mathcal{N}\left( t_{f}\right)$, is proportional to $t_f^2$, while in the
 latter, $\mathcal{N}\left( t_{f}\right)$ is proportional to $t_f$, as shown
 in equations (\ref{Ntf1+1}) and (\ref{Ntf3+1}). This occurs because in the 1+1 case
 we have a closed cavity, while the system formed by two parallel plates correspond, in fact, to an open cavity.

 The possibility of suppression of the DCE \cite{Bruno-1,Bruno-2} was not investigated,
 since we considered here always resonant cavities. It would be interesting to study a
 massless scalar field in 3+1 dimensions submitted to a RBC at one moving plate  and check
 if suppression of the DCE still occurs. We think that RBC, as well as more realistic BC,
 should be more investigated in  the DCE, whose experimental verification seems imminent
 \cite{BraggioEtAl-2005} (see also the recent proposal of experiment \cite{Squid-2009}).
 In this work we were concerned only with the regions inside the cavities, but an analysis
 involving also the outside regions,
including a discussion of the dissipative force on the moving plate and the energy balance,
 can be made and will appear elsewhere.

\noindent
{\bf Acknowledgments:} C.F. would like to thank CNPq and Faperj for a partial financial support.



\end{document}